\documentclass[%
 reprint,
superscriptaddress,
 amsmath,amssymb,
 aps, prl
]{revtex4-1}

\usepackage{amsmath}
\usepackage{graphicx}
\usepackage{caption} 
\usepackage{subcaption} 
\usepackage{cleveref}
\usepackage{float}
\usepackage{dcolumn}
\usepackage{bm}

\begin{document}

\preprint{APS/123-QED}

\title{The attosecond regime of impulsive stimulated electronic Raman excitation
}

\author{Matthew R. Ware}
\affiliation{Department of Physics, Stanford University, Stanford, California 94305, USA}
\affiliation{Stanford PULSE Institute, SLAC National Accelerator Laboratory, Menlo Park, CA 94025, USA}

\author{Philip H. Bucksbaum}
\affiliation{Department of Physics, Stanford University, Stanford, California 94305, USA}
\affiliation{Stanford PULSE Institute, SLAC National Accelerator Laboratory, Menlo Park, CA 94025, USA}
\affiliation{Department of Applied Physics, Stanford University, Stanford, California 94305, USA}

\author{James P. Cryan}
\affiliation{Stanford PULSE Institute, SLAC National Accelerator Laboratory, Menlo Park, CA 94025, USA}

\author{Daniel J. Haxton}
\affiliation{Department of Physics, University of California, Berkeley, CA 94720,
USA}
\email{djhaxton@berkeley.edu}

\date{\today}
\begin{abstract}
We have calculated the resonant and nonresonant contributions to attosecond impulsive stimulated electronic Raman  scattering~(SERS) in regions of autoionizing transitions.  
Comparison with Multiconfiguration Time-Dependent Hartree-Fock~(MCTDHF) calculations find that attosecond SERS is dominated by continuum transitions and not autoionizing resonances.  These results 
agree quantitatively with a rate equation  that includes second-order Raman and first- and second- order photoionization rates.
Such rate models can be extended to 
larger molecular systems.
Our results indicate that attosecond SERS transition probabilities may be understood in terms of two-photon generalized cross sections even in the high-intensity limit for extreme ultraviolet wavelengths.


\end{abstract}

\maketitle


The development of intense, broadband extreme ultraviolet (XUV) and x-ray pulses has spurred interest in nonlinear x-ray spectroscopic techniques with infrared and visible wavelength analogs ~\cite{mukamel_multidimensional_2013}. 
These techniques employ interactions with sequences of laser pulses in order to probe the correlations between different molecular states, and thus require extremely high intensities, which are only now being realized~\cite{bostedt_linac_2016,takahashi_attosecond_2013}.  

\begin{figure}
\includegraphics[width=0.7\linewidth]{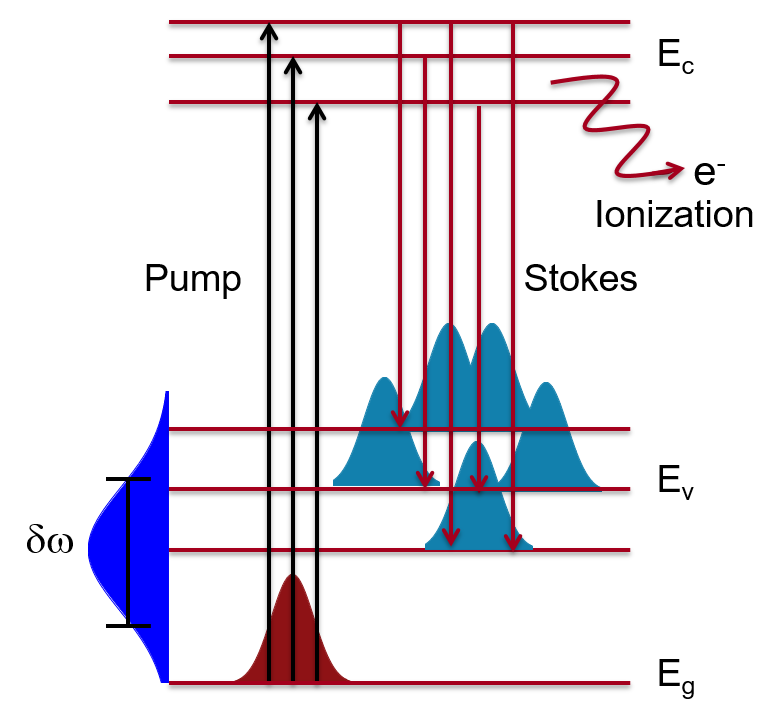}
\captionsetup{justification=raggedright,
singlelinecheck=false
}
\caption{In impulsive stimulated electronic Raman scattering, an atom or molecule simultaneously absorbs and emits a photon from an attosecond pulse having bandwidth greater than the ground-to-excited state energy, ie. $\delta \omega > E_v -E_g$. The intermediate core-ionized state, $E_c$, is either a cationic state, such as Na$^+$ $2p^{-1} 3s^1$+e$^-$, or an autoionizing state such as Na $2p^{-1}3s^14s^1$. The rate of attosecond Raman excitation is faster than the rate of auto- or direct ionization in the high intensity limit.} \label{fig:isrs}
\end{figure}

One such nonlinear technique, which has been proposed as a method for preparing valence electronic wavepackets, is impulsive stimulated electronic Raman scattering (SERS). 
The SERS technique, depicted in Fig.~\ref{fig:isrs}, begins with the excitation of an inner-shell~(or core) electron into an unoccupied state (either in the valence band or the ionization continuum) before this highly excited state can decay, a second interaction with the laser field stimulates emission of a photon and the core-vacancy is filled. 
The resulting final state can be any valence-excited state within the bandwidth of the exciting laser pulse. 
Moreover, since the initial core-orbitals are spatially localized, the resulting wavepacket should also begin localized around a given atom in a molecular system. 

In this work we focus on the exact calculation of generalized cross-sections for SERS with a single, extremely broad bandwidth, XUV laser pulse using our implementation of the Multiconfiguration Time-Dependent Hartree-Fock~(MCTDHF) method. 
We also implement a rate equation model in order to glean a physical intuition from the exact result. 
One striking observation is that once the exciting laser pulse becomes broad enough to drive SERS, the major contribution to the SERS process is due to the unstructured continuum, and the presence of auto-ionizing states has little to no effect on the generalized Raman cross-section. 

\begin{figure*}
\includegraphics[width=.9\textwidth]{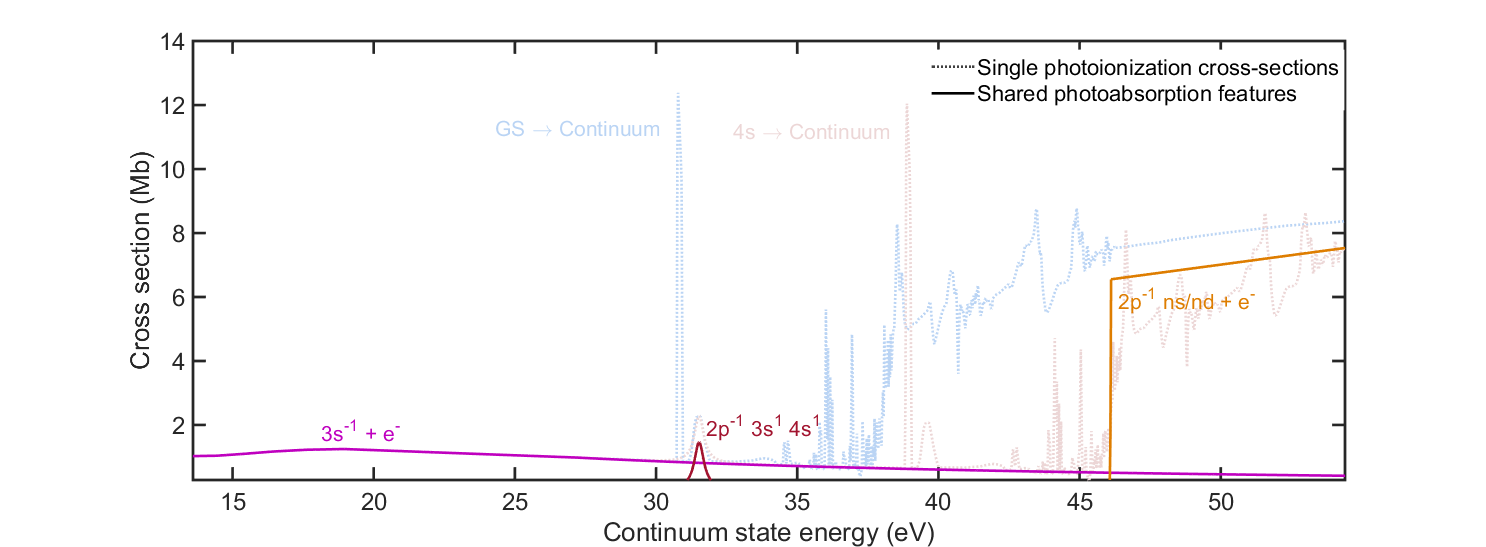}
\captionsetup{justification=raggedright,
singlelinecheck=false
}
\caption{The measured photoabsorption cross-section from the ground state of sodium as measured in Ref. \cite{wolff_photoabsorption_????} is shown in dotted blue as a function of absolute energy of the continuum states ($\sigma_{g}^{(1)}(E) = \sigma_{g}^{(1)}(\omega + E_g)$), the approximated photoabsorption cross-section from the 4s state is shown in dotted red, and the shared features between these spectra are shown in solid lines. The $3s^{-1}+e^-$ and $2p^{-1}  3s^{-1}  ns/nd+e^-$ continua are shared by the ground and the 4s states, as well as the $2p^{-1}  3s 4s$ autoionizing resonance. All of these shared structures, $\sigma_{GS-j}^{(1)}$ and $\sigma_{E-j}^{(1)}$, contribute to the Raman cross-section in Eqn.~\ref{eqn:raman_cs}.} \label{fig:photoabsorption}
\end{figure*}

The usefulness of impulsive stimulated Raman scattering for creating and probing coherent charge dynamics has already been investigated theoretically~\cite{biggs_two-dimensional_2012,biggs_watching_2013,schweigert_probing_2007,healion_simulation_2011,mukamel_multidimensional_2013}.
However these previous works focused primarily on the information content of the Raman signal and not the overall yield of the Raman excitation, i.e. the valence population transfer. 
MCTDHF has already been used to calculate population transfer for a set of narrow bandwidth laser pulses with different central energies~\cite{haxton_ultrafast_2014,li_population_2014}.
More recently, MCTDHF was used to validate a two-color optimal-control scheme in neon~\cite{greenman_laser_2015}. 
Other works which consider population transfer via SERS use model systems that focus on transitions via the autoionizing resonances~\cite{miyabe_transient_2015,picon_stimulated_2015,weninger_stimulated_2013-1}.


A simple and general model for calculating Raman population transfer is an important step towards developing the aforementioned nonlinear spectroscopies. In lieu of experimental data, MCTDHF calculations~\cite{haxton_multiconfiguration_2011,haxton_https://commons.lbl.gov/display/csd/lbnl-amo-mctdhf_????} are a leading way to confirm that any model of SERS captures the necessary physics to describe the Raman process as demonstrated by Ref.~\cite{li_population_2014}.

In this work,
we consider linearly polarized XUV pulses where $E(t) = \partial_t A(t)$ and the vector potential is given by
\begin{equation}
A(t) = \sin(\omega_0 t) \sin^2\left(\frac{\pi t}{2 \Delta t}\right) \quad 0 \le t \le 2 \Delta t,
\end{equation}
where the pulse duration, measured as the full-width at half maximum, is $\Delta t$.
In this work we consider laser pulses with FWHM duration (energy bandwidth) 1~fs~(3.25~eV), 500~as~(6.5~eV), and 250~as~(13~eV) interacting with atomic sodium. 

Briefly, the MCTDHF method~\cite{kitzler_ionization_2004,kato_time-dependent_2004,caillat_correlated_2005,nest_multiconfiguration_2005,alon_unified_2007,nest_time-dependent_2007,nest_pump_2008,kato_time-dependent_2008,kato_time-dependent_2009,kato_ionization_2009,miyagi_time-dependent_2013,sato_time-dependent_2013,sato_time-dependent_2015,sawada_implementation_2016}
solves the time-dependent Schrodinger equation using a time-dependent 
linear combination of Slater determinants, with time-dependent orbitals in the Slater determinants.  
The nonlinear working equations are obtained through
application of the Lagrangian variational principle to this wave function ansatz~\cite{broeckhove_equivalence_1988,ohta_time-dependent_2004}.

All ionization channels are implicitly included in a MCTDHF calculation 
because all the orbitals are time-dependent.  The Slater determinant list 
for the 11 electrons in sodium is
defined by 10 orbitals and full configuration interaction.  In the initial
state, four of the orbitals have $s$ symmetry and the remaining six orbitals 
comprise two $p$ shells.  The projection of angular momentum on the z axis
(the direction of linear polarization)
is conserved during the pulse, so six orbitals maintain their $m$=0 quantum
number and two orbitals each remain $m=1$ and $m=-1$.
The orbitals are represented
using a finite element discrete variable representation (FEM-DVR)~\cite{mccurdy_solving_2004,rescigno_numerical_2000} in the radial coordinate with Legendre DVR in the polar angle 
$\theta$.  We include seven DVR basis functions in $\theta$, with 18 functions
per radial finite element, five finite elements 6 bohr long followed by
four finite elements 6, 6, 9, and 15 bohr long complex-coordinate-scaled at 30 degrees.  With this basis we have converged the calculation
with respect to the primitive representation of the orbitals and the results
are gauge-invariant~\cite{simon_definition_1979}.


For comparison, we model the process of SERS using a variant of the effective three-state model described by Refs.~\cite{miyabe_transient_2015,li_population_2014,picon_stimulated_2015}. In addition to autoionizing resonances, our model also explictly includes the continuum. 
We employ a two-state rate equation model for the populations, $\vec{N}(t)=\left[N_{GS}(t), N_{4s}(t)\right]$, of the initial and final state, which may be written
%
%
\begin{eqnarray}
\frac{\vec{N}(t)}{dt} &=& M(t) \vec{N}(t) \label{eqn:rate_model} \\
M(t) &=& \left(\begin{smallmatrix}
    -{\sigma}_R^{(2)}I(t)^2 - \sum_{n=1}^{2} {\sigma}_{GS}^{(n)}I(t)^n   & {\sigma}_R^{(2)}I(t)^2  \\ {\sigma}_R^{(2)}I(t)^2  &  -{\sigma}_R^{(2)}I(t)^2 - \sum_{n=1}^{2} {\sigma}_{E}^{(n)}I(t)^n \nonumber
\end{smallmatrix} \right),
\end{eqnarray}
where ${\sigma}_R^{(2)}$ and ${\sigma}_i^{(n)}$ are the generalized non-sequential Raman cross section and $n$-photon non-sequential ionization cross sections for the state $i$.

We approximate the generalized cross sections 
using products of single photoabsorption cross sections from Ref. \cite{wolff_photoabsorption_????} as described
%
%
in the supplementary material. The 
resulting two-photon cross section is given by:
\begin{equation}
\begin{split}
{\sigma}_R^{(2)}(\omega_0,\delta \omega) = &2\pi \sum_j \int d\omega_1 \int d\omega_2 \sigma_{GS-j}^{(1)}(\omega_1) \\ &\sigma_{E-j}^{(1)}(\omega_1-\omega_2) f(\omega_1; \omega_0,\delta \omega) f(\omega_2; \omega_0,\delta \omega) \\&\delta(\omega_1-\omega_2-\omega_{GS-E}) p_j(\omega_1)
\end{split}
\label{eqn:raman_cs}
\end{equation}
\begin{equation}
\begin{split}
{\sigma}_i^{(2)}(\omega_0,\delta\omega) =&2\pi \sum_j\int_0^\infty d\omega_1 \sigma_i^{(1)}(\omega_1)f(\omega_1)p_j(\omega_1) \\ &
\int_0^\infty d\omega_2 \sigma_i^{(1)} (\omega_1+\omega_2)f(\omega_2),
\end{split}
\label{eqn:photoionization_cs}
\end{equation}
where $p_j(\omega)$ 
is the decay rates of the continuum (cation + free electron) and autoionizing states. $f(\omega;\omega_0,\delta \omega)$ is the pulse envelope for a pulse of bandwidth $\delta \omega$ centered at $\omega_0$. $\sigma_i^{(1)}(\omega)$ is the single photoionization cross section of the ground or excited state, and $\sigma_{GS-j}^{(1)}(\omega)$ are the partial cross sections for ionization of the ground state to the j$^{th}$ continuum, shown in Fig.~\ref{fig:photoabsorption}. 
We approximate the partial cross sections for ionization of the excited state to the j$^{\mbox{th}}$ continuum, $\sigma_{E-j}^{(1)}(\omega)$, by applying an energy shift to $\sigma_{GS-j}^{(1)}(\omega)$, as shown in Fig.~\ref{fig:photoabsorption}.
In the derivation of Eqns.~\ref{eqn:raman_cs}~and~\ref{eqn:photoionization_cs} we consider only non-sequential i.e. direct contributions to the Raman (photoionization) cross section, meaning the intermediate state is not populated during the Raman (ionization) transition. 
We are justified in neglecting sequential contributions due to the scaling of these contributions with the pulse duration: the probability of sequential processes scales with the square of the pulse duration, whereas for non-sequential processes the probability is independent of the pulse duration~\cite{feist_nonsequential_2008}. 

\begin{figure}
    \begin{subfigure}{\linewidth}
        \includegraphics[width=0.75\linewidth]{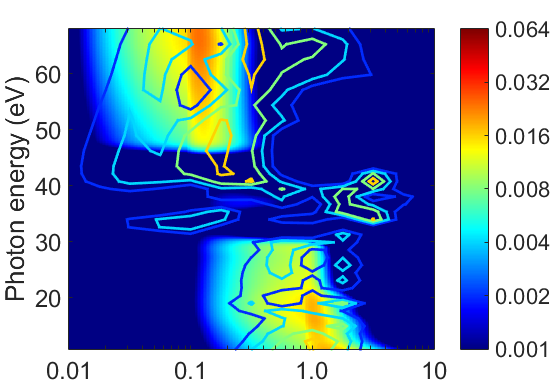}
    \end{subfigure}%
    \\
    \begin{subfigure}{\linewidth}
        \includegraphics[width=0.75\linewidth]{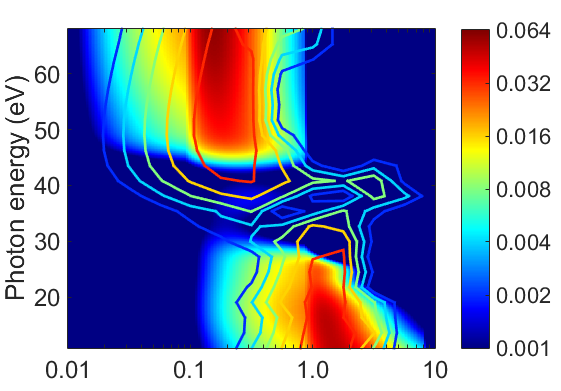}
    \end{subfigure}%
    \\
    \begin{subfigure}{\linewidth}
        \includegraphics[width=0.75\linewidth]{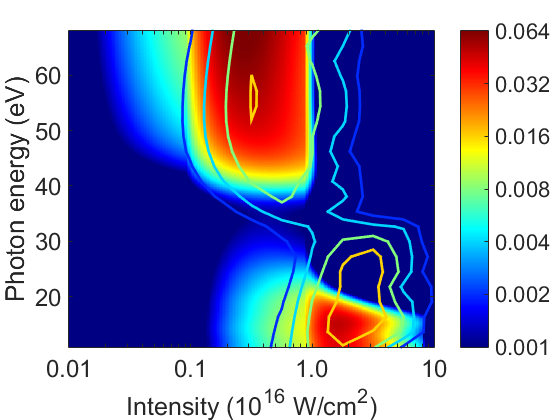}
    \end{subfigure}
    \captionsetup{justification=raggedright,
singlelinecheck=false
}
    \caption{Valence excited state~(sodium $4s$) populations, calculated using MCTDHF~(contours) and rate equation model~(colormap), following the interaction with laser pulses of different bandwidths: (top)~3.25~eV, (middle)~6.5~eV, and (bottom)~13~eV.}
    \label{fig:raman_spectra}
\end{figure}

This model finds excellent agreement with the exact MCTDHF results for pulses with bandwidths of 13~eV~(FWHM) and below, as shown in Fig.~\ref{fig:raman_spectra}. 
The rate equation model overestimates the Raman transition probability at very high energy bandwidths ($\gtrsim$13~eV) because we have ignored the photon energy dependent phase of the transition dipole moments in Eqn.~\ref{eqn:raman_cs}, which would lead to interference. 
The model also misses the fine structure at 40~eV which we attribute to a sequential Raman process through the $3s^{-1}+e^-$ continuum and the $2p^{-1}ns/nd+e^-$ continuum.
We confirm these features arise from a sequential process because the probability of these transitions scales with the square of the pulse duration, as expected for a sequential process~\cite{feist_nonsequential_2008}.  

\begin{figure}
\includegraphics[width=.8\linewidth]{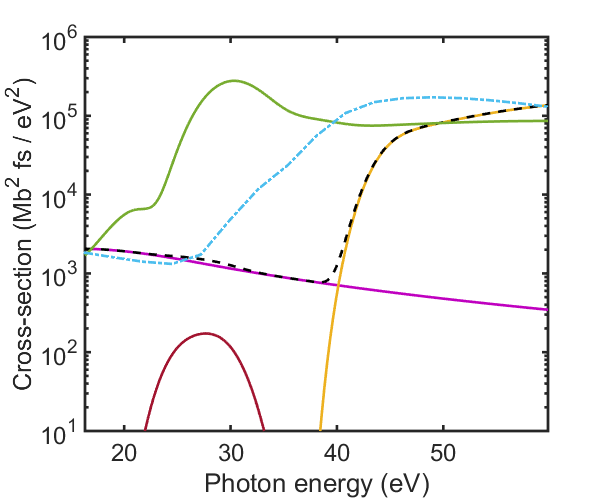}
\captionsetup{justification=raggedright,
singlelinecheck=false
}
\caption{Generalized Raman cross section from Eqn.~\ref{eqn:raman_cs} for a 500~as (6.5~eV~FWHM) XUV pulse (black line). The green curve shows the two-photon ionization cross section as a function of photon energy from Eqn.~\ref{eqn:photoionization_cs}. Partial Raman cross sections due to the $3s^{-1}$ continuum (pink), $2p^{-1}$ continuum (yellow), and the auto-ionizing states (red) are also shown. The MCTDHF result is shown in blue.} \label{fig:half_fs_2nd_cs}
\end{figure}

From the MCTDHF results in Fig. \ref{fig:raman_spectra}, we may extract 
a generalized two-photon Raman
cross section by taking a line out of the Raman probability at very low intensity. 
This line out is then scaled by the intensity and pulse width. 
The result for a 6.5~eV FWHM pulse is shown in Fig.~\ref{fig:half_fs_2nd_cs}. 
We see that the total generalized cross section for Raman transitions is of the same order of magnitude as the estimated two-photon generalized cross section for photoionization 
retrieved with Eqn.~\ref{eqn:photoionization_cs}. 
Because of this, the attosecond SERS transition rate and the dication production 
rate are comparable at the two peaks in Fig.~\ref{fig:raman_spectra}. 

Interestingly, at 500~as the MCTDHF results show no sensitivity to the autoionizing states, which lie in the energy range of 30--40~eV, as demonstrated by the monotonic increase of the generalized two-photon Raman cross section across this energy range, instead of peak structure which would be expected for an intermediate resonance. 
%
%
With our rate equation model, we are able to extract the partial cross sections for the Raman process through the continuum and autoionizing states, which are shown in Fig.~\ref{fig:half_fs_2nd_cs}. 
We find that the transition rates through the autoionizing states are an order of magnitude weaker than through the unstructured continuum. 
Moreover, the autoionizing states appear to contribute very strongly to double photoionization, despite their weak contribution to the Raman cross section. 
Therefore we find that the autoionizing states are primarily a channel for double photoionization and not a strong channel for population transfer in attosecond SERS at XUV wavelengths.

We can use our model to determine the condition where transitions through the unstructured continuum dominate autoionizing resonances.
Looking again at Eqn.~\ref{eqn:raman_cs}, the partial cross section for the continuum is approximately ${\sigma}_{R,C}^{(2)}(\omega) \sim \sigma_{GS-C}^{(1)}(\omega_0)^2~p_c(\omega_0)~\delta\omega$, and the partial cross section for autoionizing resonances is approximately ${\sigma}_{R,AI}^{(2)}(\omega_{AI}) \sim \sigma_{GS-AI}^{(1)}(\omega_{AI})^2~\Gamma^2/\omega_{GS-E}^2$. 
Therefore in the limit where
\begin{equation}
\left(\frac{\sigma_{GS-AI}^{(1)}}{\sigma_{GS-C}^{(1)}}\right)^2\ll p_c~\delta\omega \left(\frac{\omega_{GS-E}}{\Gamma}\right)^2,
\label{eqn:atto_regime}
\end{equation}
at $\omega_0 = \omega_{AI}$, the rate of population transfer
through the autoionizing resonance is much slower than the rate of population transfer
through the unstructured continuum. 
The condition described in Eqn.~\ref{eqn:atto_regime} is satisfied in the current situation, i.e. sub-fs duration XUV pulses interacting with atomic sodium, but we have also confirmed that for sub-fs duration XUV, Eqn.~\ref{eqn:atto_regime} is often fulfilled.
For x-rays, the situation is more complex due to significant Stark shifting of the states, but Eqn.~\ref{eqn:atto_regime} still allows identification of impulsive continuum transitions when the Stark shift is taken into account, for more information see Ref.~\cite{ware_red-detuned_2016}. 

This leads to a rather interesting observation, even though the excitation pulse is always in resonance with the electronic continuum, if the single photon absorption cross section is approximately constant over the pulse bandwidth, $\sigma_{i}^{(1)}(\omega)\sim c$, then Eqn.~\ref{eqn:raman_cs} becomes a Fourier-type integral similar to that found in nonresonant impulsive Raman for vibrations \cite{silberberg_quantum_2009}. 
This suggests that we can draw intuition about impulsive electronic Raman from earlier studies on impulsive vibrational Raman. 


The impulsive creation of a coherent electronic excitation in the attosecond regime provides an exciting opportunity to study electrons in atoms and molecules on their natural timescale.
The natural timescale for electron motion in atoms and molecules is extremely fast; electrons can move across molecular bonds in less than a femtosecond. 
The dynamics of electrons on this extreme timescale is at the forefront of both experimental and theoretical ultrafast many-body physics~\cite{lepine_attosecond_2014,corkum_attosecond_2007,leone_what_2014}. 
Observations of electron motion on this time scale would greatly increase our understanding of energy migration and charge transfer in molecular systems. 
One particularly interesting problem is understanding the dynamics of a coherent electronic excitation in a molecular system~\cite{kuleff_ultrafast_2014}. 
Various Raman signals have already been considered in literature as a way to probe these dynamics.

However, the accurate calculation of excited state populations resulting from Raman scattering is not nearly as mature, most notably when considering the impulsive limit, where both the pump and stokes frequencies are contained in a single laser pulse. 
To this end we have used our implementation of MCTDHF to predict the excited state populations resulting from impulsive Raman scattering with sub-fs XUV pulses in a model system, atomic sodium.
We have also developed a rate-equation based model which accurately describes the Raman process.
The relative agreement between the MCTDHF result and the rate-equation model shown in Fig.~\ref{fig:raman_spectra} and Fig.~\ref{fig:half_fs_2nd_cs} is very encouraging, since the model only uses measured ionization cross sections and hence can easily be adapted to more complicated target atoms and molecules. 
By comparison, the implementation of MCTDHF for larger systems requires large amounts of computational resources and time. 
Moreover, our rate equation model allows the investigation of how laser intensity, central frequency, bandwidth, and chirp effect the impulsive SERS process, which we will pursue in future reports. 

The rate equation model allowed us to investigate the different partial Raman cross sections, and we report that
for extremely broad bandwidth laser pulses (sub-fs in duration) the auto-ionizing states have little effect on the Raman cross sections. However, the presence of these states does contribute significantly to the loss channel (double ionization).

Matthew Ware is supported by the Stanford Graduate Fellowship. Dan Haxton is supported by the Peder Sather Grant Program. This research is further supported through the Stanford PULSE Institute 
by the U.S. Department of Energy, Office of Basic Energy Sciences: Chemical Sciences, Geosciences, and Biosciences Division.


%

\end{document}